\documentclass[aps,prl,reprint,groupedaddress]{revtex4-2}

\usepackage{graphicx}% Include figure files
\usepackage{dcolumn}% Align table columns on decimal point
\usepackage{amsmath}
\usepackage{bm}% bold math

\usepackage[dvipsnames]{xcolor}

\begin{document}

\title{Synchronization modes in bipartite oscillator networks}
\author{Pau Pom\'es}
\author{Bastian Pietras}
\altaffiliation[Current address: ]{Natural Intelligence (NISYS GmbH), Frankfurt am Main, Germany.}
\author{Ernest Montbri\'o}
\affiliation{Neuronal Dynamics Group. Department of Engineering. Universitat Pompeu Fabra, 08003 Barcelona, Catalonia, Spain.}
\date{\today}

\begin{abstract}
Collective oscillations in neuronal systems often arise from interactions between excitatory and inhibitory populations 
rather than from recurrent coupling within a single ensemble. 
Motivated by the coexistence of strongly and partially synchronized regimes in such systems, 
we study the Kuramoto–Sakaguchi model on a bipartite network.
Despite its minimal structure, the model exhibits rich collective dynamics, including both 
continuous and discontinuous transitions from full synchrony to partial synchrony (PS). 
In the PS regime, global oscillations fail to entrain one of the two populations, 
whose oscillators display quasiperiodic dynamics with an average frequency that can 
significantly deviate from that of the global field, as observed in neuronal networks.
We show that this PS state constitutes an example of self-organized quasiperiodicity, 
arising here in the bimodal Kuramoto–Sakaguchi model despite its purely linear global coupling.
\end{abstract}

%\keywords{}

\maketitle
The emergence of coherent oscillations in large networks of self-sustained oscillators
is a ubiquitous phenomenon in nature~\cite{Win80,PRK01,Str03}. In the seminal theoretical 
frameworks of Winfree and Kuramoto~\cite{Win67,Kur84}, collective synchronization 
arises from recurrent interactions within a single, globally coupled population. 
However, global rhythms may originate instead from the cross-talk between 
distinct populations ---often of different types and coupled non-reciprocally~\cite{FHL+21}.

A paradigmatic example is found in neuronal circuits, where oscillations emerge 
from the interplay between excitatory (E) and inhibitory (I) neurons~\cite{WTK+00,Boe17}. 
Models of interacting E and I spiking neurons often exhibit a strongly synchronized (S) regime, 
where both populations fire at every cycle of the global oscillation~\cite{HM03,BK03}. 
This regime is well-described by the Kuramoto model on a bipartite network~\cite{MP18}. 
However, experimental observations frequently reveal a partially synchronized (PS) regime
~\cite{WTK+00,Boe17}, 
where E-cells skip cycles of the global rhythm while inhibitory neurons fire more regularly
%%%%%%%%%%%%%%%%%%%%%%%%%%%%%%%%%%%%%%%%%%%%%%%%%%%%%%%%%%%%%%%%%%%%%%%%%%%%%%%%%%%%%
\footnote{Previous studies have reproduced PS states in E-I networks of spiking neurons
by introducing stochastic inputs or additional currents 
in the E population~\cite{BEK05,KL94,KE11,KGG14,MRP05,Boe17}.
Alternative modeling approaches have shown that PS states arise in E-I networks with random connectivity, 
noise, and synaptic delays~\cite{Bru00,BW03,BH08,Wan10}.}. 
%%%%%%%%%%%%%%%%%%%%%%%%%%%%%%%%%%%%%%%%%%%%%%%%%%%%%%%%%%%%%%%%%%%%%%%%%%%%%%%%%%%%%

To investigate the mechanisms underlying these synchronization modes, 
we analyze an extension of the paradigmatic Kuramoto-Sakaguchi (KS) 
model~\cite{SK86} in which oscillators in population $\sigma$ 
interact only with oscillators in the opposite population  $\sigma'$
\begin{equation}
\dot{\theta_i}^{\sigma}  = \omega_{\sigma} -  \frac{K_{\sigma'}}{N} \sum_{j=1}^N 
\sin(\theta^{\sigma}_i-\theta^{\sigma'}_j -\alpha_{\sigma}), \quad  i=1,\dots,N, 
\label{KS} 
\end{equation}
where $\theta_i^{\sigma}$ is the phase of the $i$th oscillator in population 
$\sigma \in \{A,B\}$. Here, $\omega_{\sigma}$, $K_{\sigma}$, and $\alpha_{\sigma}$ are natural frequencies, 
coupling strengths, and phase-lag parameters, respectively. 
Eq.~\eqref{KS} can be formally derived from an E-I network of \emph{identical}
Quadratic Integrate-and-Fire (QIF) neurons with conductance-based synapses~ 
%%%%%
\footnote{In Ref.~\cite{MP18} a two-population Kuramoto model
was derived from an E-I network with \emph{current-based} synapses, see also~\cite{CPM22}. 
This model without self-coupling (see footnote [42] in Ref.~\cite{MP18}) 
corresponds to Eqs.~\eqref{KS} with $\alpha_{A,B}=0$.}.
%%%%%
Though QIF neurons have identical intrinsic parameters,
the non-reciprocal nature of E-I interactions generically leads to mismatches in 
frequencies ($\omega_A\neq\omega_B$), coupling strengths ($K_A \neq K_B$), 
and phase-lags ($\alpha_A \neq \alpha_B$) in the reduced Eqs.~\eqref{KS}~
\footnote{See Supplemental Material at [URL] for the derivation of the bipartie KS model, 
Eqs.~\eqref{KS}, from a bipartite network of QIF neurons.}.

A substantial body of work has investigated PS states in symmetrically, fully connected 
two-population KS networks of identical oscillators
~\cite{MKB04,AMS+08,PR08,Lai09,PAA+16,MBP16,KJA17,CPM22}. Most notably,  
PS states emerging through symmetry breaking and known as \emph{chimera states} have 
attracted enormous interest; see e.g.~\cite{PA15,PR15i,BGL+20,Hau21} for reviews. 
By contrast, the bipartite KS model has received little attention.
So far  Eqs.~\eqref{KS} have been analyzed either for 
$\alpha_\sigma = 0$ ~\cite{MP18,VM09,FHL+21,PRA15,CMM18,Tho21}, 
or for $\alpha_\sigma \neq 0$ and $\omega_A=\omega_B$~\cite{QCA+21}; 
in none of these cases have PS-states been reported. 

In this Letter, we demonstrate that both S and PS states arise from cross-population interactions 
in Eqs.~\eqref{KS}. 
This simplicity allows for a transparent analysis of the underlying mechanisms 
leading to PS states. We show that these states represent a novel example of 
\emph{self-organized quasiperiodicity} (SOQ)~\cite{RP07,PR09}, 
which here emerge solely from the bipartite structure of the network 
and do not require the presence of nonlinear mean field coupling~\cite{RP07,PR09,LP19}.

In what follows, we set $\alpha_A=\alpha_B=\alpha$ without loss of generality 
\footnote{Defining $\vartheta_i^{B} =\theta_i^{B}+(\alpha_A-\alpha_B)/2$, 
the phase shift parameters in Eq.~\eqref{KS}  
become $\alpha_A=\alpha_B=\alpha$, with $\alpha=(\alpha_A+\alpha_B)/2$.
If the two population KS model has self-interactions, $\alpha_A=\alpha_B$ 
cannot be adopted without loss of generality.}.
Moreover, we consider the case $K_A=K_B=K>0$, $\alpha\in[0,\pi/2)$, and $\Delta=\omega_A-\omega_B>0$~
\footnote{See Supplemental Material at [URL] for the general case, which does not add new qualitative behavior.}.

%%%%%%%%%%%%%%%%%%%%%%%%%%%%%%%%%%%%%%%%%%%%%%%%%%%%%%%%%%%%%%%%%%%%%%%%%%%%

%%%%%%%%%%%%%%%%%%%%%%%%%%%%%%%%%%%%%%%%%%%%%%%%%%%%%%%%%%%%%%%%%%%%%%%%%%
\begin{figure}[t]
\includegraphics[width=1\columnwidth]{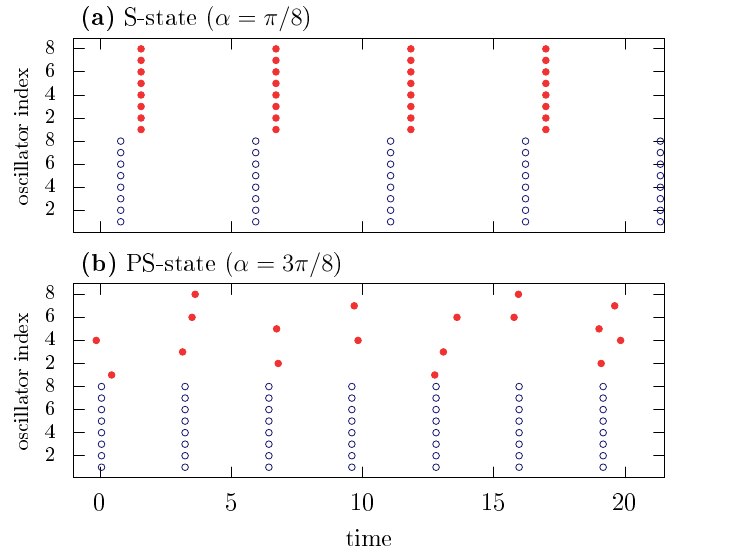}
\caption{Synchronization modes of Eqs.~\eqref{KS},
with $N=8$. Raster plots of populations A 
(open circles, Blue) and B (closed circles, Red) in the 
(a) S-state ($\alpha=\pi/8$), (b) PS-state ($\alpha=3\pi/8$).
Parameters: $K=1$, $\omega_A=1.75$, $\omega_B=0.25$, ($\bar \omega=1$, $\Delta=1.5$).}
\label{fig:Raster}
\end{figure}
%%%%%%%%%%%%%%%%%%%%%%%%%%%%%%%%%%%%%%%%%%%%%%%%%%%%%%%%%%%%%%%%%%%%%%%%%%

\emph{Numerical illustration of S- and PS-states.---}The raster plots 
in Figure~\ref{fig:Raster} illustrate the 
two synchronization modes of the KS model Eqs.~\eqref{KS}. 
The time points at which the oscillators' 
phases reach a multiple of $2\pi$ are depicted, after a transitory period of 1000 time units --- initial phases 
were randomly drawn from a uniform distribution in $(0,2\pi]$. 
In the S regime (panel a), oscillators in populations A (blue open dots) and B 
(red dots) are in-phase synchronized, and the two populations are locked to a common frequency $\Omega$. 
In this case, the network dynamics reduce to that of two 
mutually coupled oscillators, with a constant phase mismatch 
$\phi^*=\theta_i^{A}-\theta_j^{B}>0$, for all $i,j$.
In the PS regime (panel b), A-oscillators become in-phase synchronized, while B-oscillators 
remain asynchronous. Note that B-oscillators rotate non-uniformly with a time-averaged frequency 
clearly below that of A-oscillators, mirroring the PS states observed in E-I neural networks.

%%%%%%%%%%%%%%%%%%%%%%%%%%%%%%%%%%%%%%%%%%%%%%%%%%%%%%%%%%%%%%%%%%%%%%%%%%%%%%%%%%%%%%%

\begin{figure}[t]
\includegraphics[width=1\columnwidth]{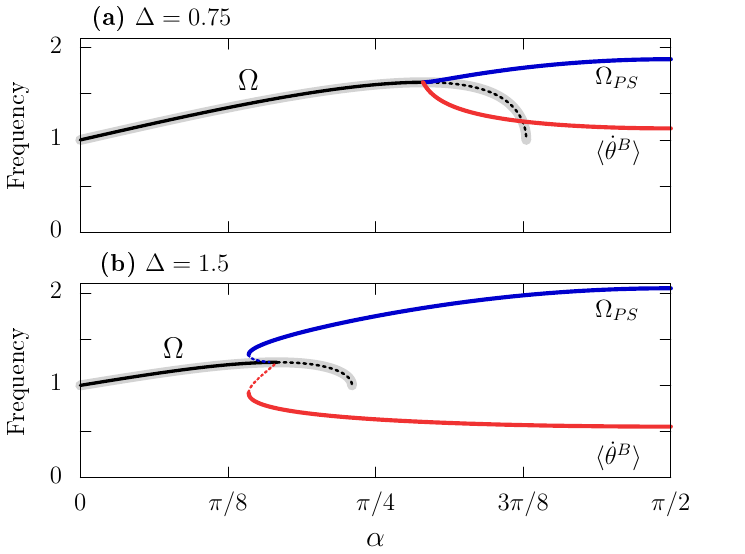}
\caption{Frequencies of the mean fields in the S 
($\Omega_{S}$: Black lines, Eq.~\eqref{OmS}) and 
PS ($\Omega_{PS}$: Blue  lines; Eq.~\eqref{OmPS})
sates, and time-averaged frequency of asynchronous oscillators in the PS state 
($\langle{\dot \theta^{B}}\rangle$: Red lines; Eq.~\eqref{OmAv}).
Grey thick line: Function $\Omega(\alpha)$, Eq.~\eqref{OmS}. 
Solid and dashed lines: Stable and unstable states, respectively. 
Parameters: $K=1$, $\bar \omega=1$, and (a) $\omega_A=1.375$, $\omega_B=0.625$;
(b) $\omega_A=1.75$, $\omega_B=0.25$.}
\label{fig:Omega}   
\end{figure}

%%%%%%%%%%%%%%%%%%%%%%%%%%%%%%%%%%%%%%%%%%%%%%%%%%%%%%%%%%%%%%%%%%%%%%%%%%%%%%%%%%%%%%%
\emph{Existence of S-states.---}We begin by analyzing Eqs.\eqref{KS} in the case of two coupled 
non-identical oscillators, $N=1$~\cite{Kur84,SSK88}.  Two possible regimes exist:
%% BP: "An" S-state instead of "a" S-state
An S-state with the dynamics shown in Fig.~\ref{fig:Raster}(a), 
and an asynchronous regime where oscillators A and B are not 
frequency-locked but undergo quasiperiodic oscillations. 
The evolution of their phase difference $\phi = \theta_1^{A}-\theta_1^{B}$ obeys 
the Adler equation
$\dot \theta_1^{A}-\dot \theta_1^{B}=\dot\phi =\Delta - 2K \cos\alpha  \sin \phi$, 
which has a saddle-node (SN) bifurcation at
\begin{equation}
\alpha_{S}=\arccos\left(\tfrac{\Delta}{2K}\right), 
\label{SN1}
\end{equation}
if $\tfrac{\Delta}{2K}\in(0,1)$. For $\alpha\in [0,\alpha_S)$,
there is a stable equilibrium ('S-state')
\begin{equation}
\phi^*=\arcsin[\Delta/(2K\cos\alpha)],
\label{phiS}
\end{equation}
with frequency

\begin{equation}
\Omega=\bar\omega + \tfrac{K}{2} \tan\alpha \sqrt{(2 \cos \alpha)^2-(\Delta/K)^2},
\label{OmS} 
\end{equation}
for $\bar \omega =(\omega_A+\omega_B)/2$. 
The function $\Omega(\alpha)$ (thick gray lines, Fig.~\ref{fig:Omega}) has a maximum,
$\Omega_M=\omega_B+K$, at
\begin{equation}
\alpha_M=\arccos \sqrt{\tfrac{\Delta}{2K} }.
\label{TC}
\end{equation}
At this critical point, $\Omega$ maximally deviates from $\omega_B$.
Next, we show that S-states with $\alpha>\alpha_M$ 
are inherently unstable in networks with $N>1$ and lead to the "disintegration" 
of the in-phase synchronized cluster of B-oscillators.

%%%%%%%%%%%%%%%%%%%%%%%%%%%%%%%%%%%%%%%%%%%%%%%%%%%%%%%%%%%%%%%%%%%%%%%%%%
%%%%%%%%%%%%%%%%%%%%%%%%%%%%%%%%%%%%%%%%%%%%%%%%%%%%%%%%%%%%%%%%%%%%%%%%%%

\emph{Stability of S-states.---} To elucidate the origin of this instability in large networks, 
we initialize Eqs.~\eqref{KS} in an S-state for some $\alpha \in (0,\alpha_S)$ such that 
$\theta^{A,B}_j=\psi_{A,B}$ for all $j=1,\dots, N$, 
with a phase difference $\psi_A-\psi_B=\phi^*$ defined by Eq.~\eqref{phiS}. 
We then study the dynamics of a single oscillator "$i$" in population B 
driven by a periodic force of frequency $\Omega(\alpha)$. 
In the thermodynamic limit ($N\to \infty$), the relative phase of the oscillator, 
$\varphi_i=\psi_A - \theta^B_i + \alpha$, evolves according to the Adler equation
~\footnote{For a given $\alpha\in(0,\alpha_S)$, Eq.~(6) has two equilibria. 
The S-state corresponds to the equilibrium $\varphi_i^*=\phi^*+\alpha$, 
for which $\theta_i^B=\psi_B$.
}
\begin{equation}
\dot \varphi_i = \Omega(\alpha) - \omega_B - K  \sin \varphi_i,
\label{Adler} 
\end{equation}
where the time-independent frequency $\Omega(\alpha)$ is given by Eq.~\eqref{OmS}.
At the critical value $\Omega(\alpha_M) = \Omega_M$, Eq.~\eqref{Adler} possesses a unique fixed point, 
$\varphi_i^*= \phi^*(\alpha_M) +\alpha_M=\pi/2$, 
which defines the boundary of the synchronization region (Arnold tongue) for the driven oscillator.
Inside the Arnold tongue ($\Omega<\Omega_M$), Eq.~\eqref{Adler} 
yields two equilibria corresponding to frequency entrainment: 
a stable one at $\varphi^*_i<\pi/2$ and an unstable one at $\varphi^*_i>\pi/2$. 

Systems with $\alpha< \alpha_M$ possess an S-state that satisfies $\phi^*(\alpha)< \phi^*(\alpha_M)$
because the phase difference in Eq.~\eqref{phiS} is a strictly increasing function of $\alpha$ 
(see Fig.~4, panels c and d). 
This S-state corresponds to a fixed point of Eq.~\eqref{Adler} where 
$\varphi^*_i=\phi^*(\alpha)+\alpha <\phi^*(\alpha_M) +\alpha_M =\pi/2$, and is therefore stable. 
In contrast, systems with $\alpha >\alpha_M$ exhibit an S-state with $\phi^*(\alpha) > \phi^*(\alpha_M)$, 
which corresponds to a fixed point of Eq.~\eqref{Adler} with $\varphi^*_i>\pi/2$, and it is unstable.

\begin{figure}[t]
\includegraphics[width=.85\columnwidth]{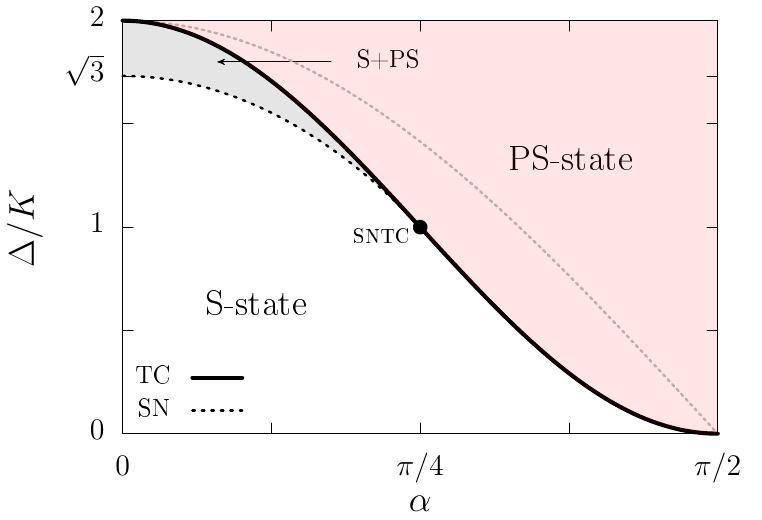}
\caption{Phase diagram of the bipartite KS model.  
Dotted thin line: SN boundary, Eq.~\eqref{SN1}.
Black solid line: Instability boundary (TC bifurcation in the OA equations),
Eq.~\eqref{TC}.
Dashed line: SN bifurcation, Eq.~\eqref{SN2}. 
Symbol "$\bullet$" : Saddle-Node-Transcritical (SNTC) point $(\pi/4,1)$. 
White region (S-state): S-state is the only stable attractor. 
Red region (PS-state): PS-State is the only stable attractor. 
Gray region (S+P): Coexistence of stable S- and PS-states.
}
\label{fig:PhD}
\end{figure}

This asymptotic analysis is corroborated by a full linear stability analysis of Eqs.~\eqref{KS}, which shows 
that S-states are unstable for $\alpha>\alpha_M$ not only in very large networks, but 
\emph{in networks of any finite size $N>1$} 
%%%%%%%%%%
\footnote{
For $N=1$, the linear stability analysis of the S-state of Eqs.~\eqref{KS} gives one zero eigenvalue, 
and a (real) eigenvalue $\mu= d \dot\phi/d\phi |_{\phi=\phi^*} = -K \sqrt{(2 \cos\alpha)^2-( \Delta/K)^2}$;
note that setting $\mu=0$ yields Eq.~\eqref{SN1}.
For $N>1$, there are two additional sets of $(N-1)$-times degenerated eigenvalues 
$\mu_A= \tfrac{1}{2}(\mu - \Delta \tan \alpha)$ and $\mu_B=\tfrac{1}{2}(\mu + \Delta \tan \alpha)$,
which describe the internal stability of the in-phase synchronized oscillators in populations 
$A$ and $B$, respectively. For $\Delta>0$ and $\alpha\in (0,\pi/2)$, 
the eigenvalues satisfy $\mu_A<\mu<\mu_B$, and hence the first potentially unstable mode is $\mu_B$.
Setting $\mu_B=0$ gives the instability boundary Eq.~\eqref{TC}.
}.
%%%%%%%%
This instability leads to the disintegration of the in-phase
synchronized cluster of $B$ oscillators. Its boundary, Eq.~\eqref{TC}, is depicted in 
the phase diagram in Fig.~\ref{fig:PhD}. 
Additionally, the synchronization frequency $\Omega$ is plotted in Fig.~\ref{fig:Omega}, 
where solid and dashed black lines indicate whether the S-state is stable or unstable, respectively.

%%%%%%%%%%%%%%%%%%%%%%%%%%%%%%%%%%%%%%%%%%%%%%%%%%%%%%%%%%%%%%%%%%%%%%%%%%
%%%%%%%%%%%%%%%%%%%%%%%%%%%%%%%%%%%%%%%%%%%%%%%%%%%%%%%%%%%%%%%%%%%%%%%%%%

%%%%%%%%%%%%%%%%%%%%%%%%%%%%%%%%%%%%%%%%%%%%%%%%%%%%%%%%%%%%%%%%%%%%%%%%%%%%%
%%%%%%%%%%%%%%%%%%%%%%%%%%%%%%%%%%%%%%%%%%%%%%%%%%%%%%%%%%%%%%%%%%%%%%%%%%%%%
%%%%%%%%%%%%%%%%%%%%%%%%%%%%%%%%%%%%%%%%%%%%%%%%%%%%%%%%%%%%%%%%%%%%%%%%%%%%%

\emph{Ott-Antonsen (OA) equations.---} Next, we demonstrate that 
beyond the critical point $\alpha=\alpha_M$, a stable PS-state exists 
characterized by a mean field of frequency $\Omega_{PS}>\Omega_M$ that does not entrain 
the oscillators of population B. 
To investigate PS-states, we analyze the thermodynamic limit of Eqs.~\eqref{KS} 
by means of the OA ansatz~\cite{OA08}.
Here we succinctly describe the basic steps to obtain low-dimensional equations describing 
the dynamics of Eqs.~\eqref{KS} in the so-called OA manifold~\cite{PR08,WS93}. Our derivation 
trivially follows that of Refs.~\cite{AMS+08,Lai09}.

We assume the existence of two probability densities $f_\sigma(\theta_\sigma,t)$ that satisfy continuity equations 
for each population, where
\begin{equation}
v_\sigma(\theta_\sigma,t) = \omega_\sigma + \frac{K}{2i} \left[ z_{\sigma'} e^{-i(\theta_{\sigma}-\alpha)}
-\bar{z}_{\sigma'} e^{i(\theta_{\sigma}-\alpha)}\right]
\nonumber 
\label{v}
\end{equation}
corresponds to the continuum limit of Eqs.~\eqref{KS} (the overbar denotes the complex conjugate), and
\begin{equation}
z_{\sigma}=r_\sigma e^{i\psi_\sigma} =
\int_0^{2\pi} e^{i\theta_\sigma} f_\sigma(\theta_\sigma,t)~ d\theta_\sigma,
\label{KOP}
\end{equation}
are the complex Kuramoto order parameters. The OA is the expansion of 
$f_{\sigma}(\theta_\sigma)$ in the Fourier series~\cite{OA08} 
\begin{equation}
f_{\sigma} (\theta_\sigma,t)= \frac{1}{2\pi}\left(1+\sum_{n=1}^{\infty} 
\left[a_\sigma(t) e^{i\theta_\sigma}\right]^n +\text{c.c.}\right), 
\label{PK}
\end{equation}
where "c.c." denotes the complex conjugate of the preceding term.    
Substituting Eq.~\eqref{PK} into the continuity equation, 
$\partial_t f_\sigma = - \partial_\theta (v_\sigma f_\sigma)$,
and using the identity $z_\sigma = \bar{a}_{\sigma}$, 
yields a system of ordinary differential equations 
\begin{equation}
\dot{z}_\sigma = i \omega_\sigma z_\sigma + \tfrac{K}{2} 
		\left[ e^{i\alpha} z_{\sigma'} - e^{- i\alpha} \bar{z}_{\sigma'} z_\sigma^2\right].
\label{OA0} 
\end{equation}
In polar coordinates, the system reduces to a three-dimensional system
%%%%%%%%%%%%%%%
%
\begin{eqnarray}
\dot r_A &=&\tfrac{K}{2} r_B (1-r_A^2) \cos(\phi-\alpha), \label{r1Eq}\\
\dot r_B &=&\tfrac{K}{2}  r_A (1-r_B^2)  \cos(\phi+\alpha), \label{r2Eq}\\
\dot \phi &=& \Delta - K r_A \tfrac{1+r_B^2}{2r_B}  \sin(\phi + \alpha)  \nonumber\\ 
&  &- K r_B  \tfrac{1+r_A^2}{2r_A} \sin(\phi - \alpha).\label{phiEq}
\end{eqnarray}
where $\phi=\psi_A-\psi_B$. These equations have a fixed point $r_A=r_B=1$ and $\phi=\phi^*$, corresponding to the 
synchronized S-state analyzed above.  
The Jacobian matrix corresponding to the S-state is triangular with three real eigenvalues $\mu,\mu_{A},\mu_{B}$,
which coincide with the eigenvalues 
determining the stability of the S-state in the finite size system, see Ref.~[39]. 

%%%%%%%%%%%%%%%%%%%%%%%%%%%%%%%%%%%%%%%%%%%%%%%%%%%%%%%%%%%%%%%%%%%%%%%%%%%%%%%%%%%%%%%%%%%%%%%%%%

%%%%%%%%%%%%%%%%%%%%%%%%%%%%%%%%%%%%%%%%%%%%%%%%%%%%
\begin{figure}[t]
\includegraphics[width=1\columnwidth]{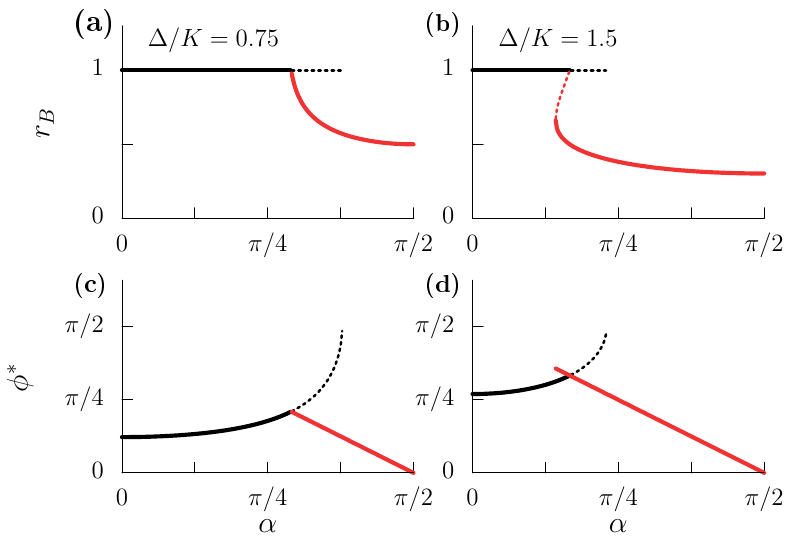}
\caption{Bifurcation diagrams of Eqs.~(\ref{r1Eq}-\ref{phiEq}), for $r_A=1$. 
Black: S-state, $r_B=1$ and $\phi^*$ given by Eq.~\eqref{phiS}. 
Red: PS-states, $r_B$ given by Eq.~\eqref{rPS}, and $\phi^*$ by Eq.~\eqref{phiPS}.
Solid and dashed lines denote stable and unstable states, respectively. 
Parameters: (a,c) $\Delta/K=0.75$; (b,d) $\Delta/K=1.5$. 
}
\label{fig:BifD}
\end{figure}
%%%%%%%%%%%%%%%%%%%%%%%%%%%%%%%%%%%%%%%%%%%%%%%%%%%%

Eqs.~(\ref{r1Eq}-\ref{phiEq}) have another type of fixed points corresponding to PS-states.
Setting $r_A=1$, the fixed point equation $\dot r_B=0$ yields a phase mismatch
\footnote{
An additional equilibrium of Eqs.~(\ref{r1Eq}-\ref{phiEq}) with $r_B=1$ and $r_A\leq1$ exists in parameter regions 
not considered in the main text [37].}
\begin{equation}
\phi^*= \pi/2-\alpha,
\label{phiPS}
\end{equation}
which depends only  on the phase lag parameter $\alpha$.
Setting $ \dot\phi=0$, we find two roots of the order parameter 
\begin{equation}
r_{B}= \left( \Delta/K  \pm  \sqrt{(\Delta/K)^2  -  4 \cos^2\alpha +1 } \right)^{-1}.
\label{rPS}
\end{equation}
Note that for $\alpha=\alpha_M$ the order parameter reaches $r_B=1$.
Hence, Eq.~\eqref{TC} is the locus of a transcritical (TC) bifurcation 
where the S and PS states exchange stability.
%%%%%%%%%%%%%%%%%%%%%%%%%%%%%%%%%%%%%%%%%%%%%%%%%%%%
\footnote{
The linear stability analysis of the PS-state 
of Eqs.~(\ref{r1Eq}-\ref{phiEq}) yields three eigenvalues: 
$\lambda=-K r_{B} \sin(2\alpha)$, and
$$\lambda_{\pm}=\lambda/2 \pm \sqrt{\lambda^2/4
+K  \left( K- r_B \Delta  \right) 
\left(1-  r_B^2 \right) /(2 r_{B}^2) },$$
where $r_B$ are given by Eq.\eqref{rPS}. The eigenvalues satisfy
$\lambda=\lambda_+ + \lambda_-<0$, and hence the only potential (stationary) 
instability is through $\lambda_+$. Note that $\lambda_+ = 0$ when ($i$) $r_B=1$ 
(TC bif.) and when ($ii$) $r_B=K/\Delta$ (SN bif.).
}
%%%%%%%%%%%%%%%%%%%%%%%%%%%%%%%%%%%%%%%%%%%%%%%%%%%%%%%%%%%%%%%%%%%%%%%%%%%%%%%%%%%%%%
In addition, Eq.~\eqref{rPS} shows that PS states are born with 
$r_B = K/\Delta \leq 1$ in a SN bifurcation at
\begin{equation}
\alpha_{PS}=\arccos\left(\tfrac{1}{2} \sqrt{ 1 + (\Delta/K)^2 }  \right),
\label{SN2}
\end{equation}
for $\Delta/K\in(1,\sqrt{3})$. 
The SN and TC bifurcations meet at a 
Saddle-Node-Transcritical (SNTC) bifurcation point when 
$r_B=K/\Delta = 1$, for $\alpha=\pi/4$. 
The unfolding of this codimension-two point 
reveals two scenarios --- see Fig.~\ref{fig:PhD}. Specifically, 
the bifurcation diagrams in Fig.~\ref{fig:BifD} show that for $\Delta/K\in (0,1)$ 
the PS-state bifurcates continuously from the S-state,
whereas for $\Delta/K\in (1,\sqrt{3})$ the transitions S $\leftrightarrow$ PS 
are discontinuous and hysteretic.
%%%%%%%%%%%%%%%%%%%%%%%%%%%%%%%%%%%%%%%%%%%%%%%%%%%%%%%%%%%%%%%%%%%%%%%%%%%%%%%%%%%%%%%%

\emph{Self-Organized Quasiperiodicity (SOQ) in Kuramoto-Sakaguchi networks.---}
Our last result concerns SOQ, a remarkable self-organization mechanism proposed to explain the emergence of partially synchronized states in \emph{single} 
populations of identical oscillators interacting 
through \emph{nonlinear} mean-field coupling~\cite{RP07,PR09}.

We now show that the same mechanism operates in the PS states of Eqs.~\eqref{KS}, 
despite the fact that the KS model depends only \emph{linearly} 
on the mean field. 
To see this, we assume $\theta_i^A=\psi_A$ for all $i$, 
and rewrite the evolution equation of the B-oscillators
in terms of their own mean-field phase $\psi_B$:
\begin{equation}
\dot \theta^B_i = \omega_B  - K \sin(\theta_i^B - \psi_B - \beta(\phi) ),
\label{SOQ}
\end{equation}
where $\beta (\phi)= \phi + \alpha$. 
The stability of the in-phase synchronized state of B-oscillators is determined by 
$d \dot\theta_i^B/d\theta_i^B =- K \cos \beta(\phi)$. Notably, using Eq.~\eqref{phiPS}, 
we find $d \dot\theta_i^B/d\theta_i^B =  0$, 
so that the oscillators are neither attracted to nor repelled from their mean-field phase $\psi_B$. 
In other words, the system operates at the critical boundary between synchrony 
and incoherence by self-adjusting the phase difference between the order parameters, Eq.~\eqref{phiPS}. 
The phase difference $\phi$ is a degree of freedom that is not available in single populations of identical oscillators, 
which require additional structure (e.g., nonlinear coupling) to sustain partial synchrony~\cite{RP07,PR09}.

Finally, we evaluate the average frequency of the oscillators in the B population and show 
that it can be made arbitrarily smaller than that of the mean field, in agreement with E–I neural oscillations. 
Using Eq.~\eqref{OA0}, we find the mean-field frequency
\begin{equation}
\Omega_{PS}= \omega_B + K (1+ r_{B}^2)/(2 r_B).
\label{OmPS}
\end{equation}
where $r_B$ is given by Eq.\eqref{rPS}. 
Note that this frequency coincides with $\Omega_M$ for $r_B=1$ 
and increases (without bound) as $r_B \to 0$, see Figs.~\ref{fig:Omega} and~\ref{fig:BifD}. 
Therefore, $\Omega_{PS} \geq \Omega_M$, so the mean-field frequency 
cannot entrain the oscillators in population B. 
Accordingly, these oscillators exhibit quasiperiodic dynamics with time-averaged frequency
\begin{equation}
\langle{\dot \theta^{B}}\rangle = \Omega_{PS} -  \sqrt{(\Omega_{PS}-\omega_B)^2-   K^2},
\label{OmAv}
\end{equation}
obtained by integrating Eq.~\eqref{Adler} with $\Omega=\Omega_{PS}$
over one rotation period. 
For $r_B=1$, $\langle \dot \theta^{B} \rangle=\Omega_{PS}$, 
while $\langle \dot \theta^{B} \rangle$ decreases as $r_B \to 0$, 
asymptotically approaching $\omega_B$, see Figs.~\ref{fig:Omega} and~\ref{fig:BifD}. 
Hence, the mean-field frequency Eq.~\eqref{OmPS} 
can differ significantly from that of the B oscillators, 
in agreement with experimental and computational studies 
in neuroscience (see, e.g.,~\cite{Boe17,BH08,Wan10} and references therein).

\emph{Conclusions.---} 
We investigated the coexistence of full and partial synchrony in E–I neural networks, 
using the Kuramoto–Sakaguchi (KS) model on a bipartite network. This canonical model, which can be formally 
derived from a network of Quadratic Integrate and Fire neurons [15], is highly amenable to analysis.  
Beyond the expected synchronized (S) state, the bipartite KS 
model exhibits a partially synchronized (PS) state that reproduces key phenomenology 
observed in E–I networks: namely, fast collective oscillations coexisting with slow, 
non-periodic individual oscillator dynamics within one population (typically the excitatory one).
However, these dynamics are quasiperiodic and lack the high irregularity characteristic of actual neuronal recordings 
or models incorporating quenched disorder and stochastic inputs~\cite{BEK05,KL94,KE11,KGG14,MRP05,BH08,Wan10}.

Finally, we found that PS in the bipartite KS model
displays a nontrivial self-organizing property: the order parameters of the two populations remain phase locked at 
a precise phase difference, such that oscillators in one population are 
neither attracted to nor repelled from their mean-field phase. 
We showed that this self-organized critical state constitutes a novel form of 
self-organized quasiperiodicity~\cite{RP07}. Yet, in contrast to the original SOQ mechanism--- introduced to explain PS in single populations of identical oscillators ---
the phenomenon observed here takes place in a system with purely linear coupling 
and is enabled by the additional degree of freedom provided by the bipartite 
structure of the network. Our results suggest that this mechanism may also operate 
in more general two-population KS networks with recurrent coupling and could 
underlie the emergence of chimera states in such systems~\cite{AMS+08,PR08,Lai09,PAA+16,MBP16,KJA17}.

\begin{acknowledgments}
The authors thank Pau Clusella and Iván León for helpful conversations. 
This work is part of Maria de Maeztu Units of Excellence Programme CEX2021-001195-M, 
funded by MICIU/AEI /10.13039/501100011033.
PP and EM acknowledge support by the Agencia Estatal de Investigaci\'on 
under the Project No.~PID2019-109918GB-I00, and by the Generalitat de Catalunya,
grant 2021 SGR0 1522 646. BP acknowledges support from the European Union’s Horizon 2020 research 
and innovation program under the Marie Skłodowska-Curie grant agreement 101032806.
\end{acknowledgments}

%\bibliography{bibliografia}

\end{document}